\documentclass[showpacs, preprintnumbers, amsmath, amssymb, floatfix, pre, twocolumn]{revtex4-2}
\usepackage{graphicx}
\usepackage{dcolumn}
\usepackage{bm}
\usepackage{amsmath,amsthm,amssymb,ascmac}
\usepackage{color}
\usepackage{subcaption}

\newcommand{\be}{\beta}
\newcommand{\ka}{\kappa}
\newcommand{\aeq}{&=}

\newcommand{\aeqd}{&:=}

\newcommand{\aeqle}{& \le}
\newcommand{\aeqge}{& \ge}
\newcommand{\aeqap}{& \approx}
\newcommand{\bra}{\langle}
\newcommand{\dbra}{\langle  \hspace{-0.5 mm} \langle}
\newcommand{\ket}{\rangle}

\newcommand{\mP}{\mathcal{P}}
\newcommand{\mQ}{\mathcal{Q}}
\newcommand{\mR}{\mathcal{R}}
\newcommand{\mB}{\mathcal{B}}
\newcommand{\mC}{\mathcal{C}}
\newcommand{\mD}{\mathcal{D}}
\newcommand{\mL}{\mathcal{L}}
\newcommand{\mU}{\mathcal{U}}

\newcommand{\mJ}{\mathcal{J}}
\newcommand{\mI}{\mathcal{I}}
\newcommand{\mH}{\mathcal{H}}
\newcommand{\mM}{\mathcal{M}}
\newcommand{\mN}{\mathcal{N}}
\newcommand{\fA}{\mathfrak{A}}
\newcommand{\fB}{\mathfrak{B}}
\newcommand{\tr}{{\rm{Tr}}}
\newcommand{\dg}{^\dagger}
\newcommand{\tl}{\tilde}
\newcommand{\defe}{:=}

\newcommand{\ep}{\varepsilon}
\newcommand{\ga}{\gamma}
\newcommand{\Ga}{\Gamma}
\newcommand{\al}{\alpha}
\newcommand{\bu}{\bullet}
\newcommand{\sig}{\sigma}
\newcommand{\f}{\frac}
\newcommand{\half}{\frac{1}{2}}
\newcommand{\pr}{\prime}
\newcommand{\dl}{\delta}
\newcommand{\Dl}{\Delta}
\newcommand{\lm}{\lambda}
\newcommand{\Lm}{\Lambda}

\newcommand{\om}{\omega}
\newcommand{\Om}{\Omega}

\newcommand{\com}{, \quad}
\newcommand{\spa}{\quad}
\newcommand{\la}{\label}
\newcommand{\ke}[1]{ \vert #1 \rangle }
\newcommand{\br}[1]{ \langle #1 \vert }
\newcommand{\abs}[1]{\vert #1 \vert }
\newcommand{\no}{\nonumber}

\newcommand{\re}[1]{(\ref{#1})}
\newcommand{\res}[1]{\S \ref{#1}}

\newcommand{\hs}{\hspace}

\newcommand{\bv}[1]{\big \vert_{#1}}
\newcommand{\Bv}[1]{\Big \vert_{#1}}

\newcommand{\RM}[1]{{\rm{#1}}}
\newcommand{\p}{\partial}
\newcommand{\co}[1]{``{#1}''}
\newcommand{\dbr}[1]{ \langle  \hspace{-0.5 mm} \langle #1 \vert }
\newcommand{\dke}[1]{\vert #1 \rangle \hspace{-0.5 mm} \rangle }

\newcommand{\bea}[1]{\begin{align} 
#1
\end{align}}
\newcommand{\rd}[1]{\textcolor{black}{#1}}

\begin{document}
\title{SLD Fisher information for kinetic uncertainty relations}

\author{Satoshi Nakajima}
\email{subarusatosi@gmail.com}
\author{Yasuhiro Utsumi}

\affiliation{
Department of Physics Engineering, Faculty of Engineering, Mie University, 
Tsu, Mie 514-8507, Japan}
\date{\today}

\begin{abstract}

We investigate a symmetric logarithmic derivative (SLD) Fisher information for kinetic uncertainty relations (KURs)
 of open quantum systems described by the GKSL quantum master equation with and without the detailed balance condition. 
In a quantum kinetic uncertainty relation derived by Vu and Saito [Phys. Rev. Lett. \textbf{128}, 140602 (2022)], 
the Fisher information of probability of quantum trajectory with a time-rescaling parameter plays an essential role. 
This Fisher information is upper bounded by the SLD Fisher information. 
For a finite time and arbitrary initial state, we derive a concise expression of the SLD Fisher information, 
which is a double time integral and can be calculated by solving coupled first-order differential equations. 
We also derive a simple lower bound of the Fisher information of quantum trajectory. 
We point out that the SLD Fisher information also appears in the speed limit based on the Mandelstam-Tamm relation 
by Hasegawa [\rd{Nat. Commun. \textbf{14}, 2828 (2023)}].  
When the jump operators connect eigenstates of the system Hamiltonian, we show that 
the Bures angle in the interaction picture  is upper bounded by the square root of the dynamical activity at short times, which contrasts with the classical counterpart.

\end{abstract}

\maketitle

\section{Introduction} 

In recent years, universal relations that characterize the fluctuations of nonequilibrium systems have been intensively investigated.
A primary class of inequalities is the thermodynamic uncertainty relation (TUR) 
\cite{TUR1, TUR2, TUR3, TUR4, TUR5, TUR6, TUR7, Vo, Yoshimura, Vu23}.
A similar relation, the kinetic uncertainty relation (KUR)
imposes another upper bound on the precision of generic counting observables in terms of the dynamical activity
\cite{KUR1, KUR2, KUR3}.

Quantum coherence plays an essential role in a broad class of thermodynamics.
Concerning the TUR and KUR originally derived for classical stochastic systems, 
it has been shown through specific examples that these relations can be violated in the quantum realm
\cite{TURB1, TURB2, TURB3, TURB4, TURB5, TURB6}.
Several quantum bounds accounting for the quantum coherence have been derived \cite{QTUR1, QTUR2, QTUR3, QTUR4, QTUR5, QKUR1}. 
\rd{For open quantum systems described by the Gorini-Kossakowski-Sudarshan-Lindblad (GKSL)  equation, Ref.\cite{QTUR1} derived a TUR using
the large deviation statistics. 
For the nonequilibrium steady states, Ref.\cite{QTUR2} derived a TUR by analyzing the total system. 
Reference \cite{QTUR5} studied periodically driven heat engines described by the quantum master equation and 
derived a TUR in the slow driving. 
For systems described by the GKSL equation 
with time-independent Hamiltonian and jump operators, Hesegawa \cite{QTUR3, QTUR4, QKUR1} derived a quantum KUR
\bea{
\f{\tau^2(\p_\tau \bra \Phi \ket)^2}{\RM{var}[\Phi]} \le \mI \la{KUR_H}
}
using the Cram\'{e}r-Rao inequality.} 
Here, $\Phi$ is a time-integrated counting observable of the system and $\RM{var}[\Phi]$ is its variance. 
$\mI$ is the symmetric logarithmic derivative (SLD) Fisher information:
\bea{
\mI \aeqd 4\Big[\p_{\theta_1}\p_{\theta_2}\mC(\theta_1,\theta_2) \no\\
&\spa-\p_{\theta_1}\mC(\theta_1,\theta_2)\p_{\theta_2}\mC(\theta_1,\theta_2) \Big]\Bv{\theta_1=0=\theta_2} \la{def_FI} ,\\
\mC(\theta_1,\theta_2) \aeqd \tr_S \rho^{\theta_1,\theta_2}(\tau).
}
Here,  $\rho^{\theta_1,\theta_2}(t)$ is the solution of the two-sided GKSL equation (see \re{two_2}) \cite{SLD}. 
$\tr_S$ is the trace of the system. 
\rd{Using this, Hasegawa estimated $\mI$ in the long-time region \cite{QTUR3}. 
For same system,} Vu and Saito \cite{Vu22} derived a quantum KUR
\bea{
\f{\tau^2(\p_\tau \bra \Phi \ket)^2}{\RM{var}[\Phi]} \le F. \la{KUR_Vu}
}
Here, $F$ is the Fisher information of probability of quantum trajectory with a time-rescaling parameter (see \re{def_theta}).
The Fisher information is the sum of the dynamical activity and the quantum correction $\mQ$ ($\mQ_2$ in Ref.\cite{Vu22})
which vanishes for the classical case. 
The calculation of $F$ is not light since one has to sum over contributions from a huge number of trajectories. 
Therefore, upper and lower bounds are useful in practical calculations. 
Vu and Saito demonstrated that $F$ \rd{is upper bounded by $\mI$}.
In this paper, we derive a concise  expression of the upper bound of $\mQ$ for a finite time and arbitrary initial state. 
We also derive a simple expression of a lower bound of $\mQ$.

Another class of inequalities is the speed limit of state transformation. 
\rd{For closed quantum systems, since 1945, the Mandelstam-Tamm relation \cite{Mandelstam, review_MT} $\int_0^\tau dt \ \Dl E \ge D$ has been known 
(In this paper, we set $\hbar=1$). 
 $\Dl E$ is the energy fluctuation and $D$ is the Bures angle (see \re{def_Bures}) between the initial and final states.
Recently, even in classical systems, it turns out that there exist speed limits expressed in terms of the distance between states \cite{Shiraishi18}. 
Shiraishi \textit{et al}. \cite{Shiraishi18} demonstrated that 
\bea{
\sqrt{\f{A(\tau)\sig}{2}}\aeqge \half l(p(0),p(\tau)) \la{SFS}
}
for a system described by a classical master equation 
$\f{d}{dt}p_n(t) =\sum_m W_{nm}p_m(t)$ satisfying the local detailed balance condition.
Here, $l(p(0),p(\tau))\defe \sum_n \abs{p_n(0) - p_n(\tau)} $ is $L_1$-norm,  $\sig$ is the total entropy production, 
and $A(\tau) \defe \int_0^\tau dt \ \sum_{n \ne m} W_{nm}p_m$ is the dynamical activity. 
A similar relation 
\bea{
A(\tau) \ge \half l(p(0),p(\tau)) \la{SL_A} 
}
have been known \cite{Vo}. 
Quantum extensions of \re{SFS} for the open quantum systems described by the GKSL equation have been researched 
\cite{Funo19, Vu21, Vu21a, Nakajima22, Vu23}.
However, the quantum extension of \re{SL_A} has been less investigated}. 
For such open quantum systems, Hasegawa \cite{H22} derived a KUR by exploiting the Mandelstam-Tamm speed limit. 
In this KUR,  a dynamical activity-like quantity appeared instead of $F$.
We point out  that this quantity equals the SLD Fisher information when Hamiltonian and jump operators are time-independent. 
Using our expressions, we derive a quantum speed limit described by the dynamical activity
when the jump operators connect the eigenstates of the system Hamiltonian.  
\rd{Our speed limit can be regarded as a quantum extension of \re{SL_A}.} 

The structure of the paper is as follows. 
First, we explain the Fisher information $F$ (\res{s_Vu}).
In \res{s_Fisher}, we show that the SLD Fisher information is a sum of the dynamical activity and a quantum correction, 
which is the upper bound of $\mQ$.
In \res{s_Numerical}, we study  $\mQ$ and its upper and lower bounds numerically in a two-level system. 
Next, we study a speed limit (\res{s_speed}). 
In \res{s_summary}, we summarize this paper. 
In Appendix \ref{A_SLD}, we explain the SLD Fisher information and the two-sided GKSL equation. 
In Appendix \ref{A_B}, we derive \re{def_I}. 
In Appendix \ref{A_correction}, we analyze the upper bound of $\mQ$ in the long-time region. 
In Appendix \ref{s_Lower}, we derive the lower bound of $\mQ$. 
In Appendix \ref{A_D}, we derive \re{dot_Q} \rd{and \re{dot_Q_-}}.
In Appendix \ref{A_Hasegawa}, we review Hasegawa's method and results.
Appendix \ref{A_QD} is for the detailed calculations for \res{s_speed}.

\section{Fisher information for quantum trajectories}  \la{s_Vu}

Here, we summarize techniques introduced in Ref.\cite{Vu22}. 
The GKSL equation is given by
\bea{
\f{d\rho(t)}{dt} \aeq \mL(t)\rho(t), \la{QME} \\
\mL(t)\bu \aeqd  -i[H_S(t), \bu] 
+\sum_k \Big[L_k(t) \bu L_k(t) \dg  \no\\
&\spa- \half \{  L_k(t) \dg L_k(t) ,\bu \} \Big].
}
Here, $[A,B]\defe AB-BA$,  $\{A, B \}\defe AB+BA$, and $\bu$ is an arbitrary linear operator of the system.
$H_S$ is the system Hamiltonian and $\{ L_k \}$ are jump operators. 
The following discussion does not require the detailed balance condition. 

A quantum trajectory is specified by a list of tuples $\Ga \defe \{ (t_1,k_1), (t_2,k_2), \cdots, (t_\mN,k_\mN)\}$. 
Here, $t_\al$ is the time of the $\al$-th jump by a jump operator $L_{k_\al}$. 
The probability of quantum trajectory is given by \cite{Vu22}
\bea{
\mP^\theta(\Ga) \aeq \tr_S[M^\theta(\Ga)\rho(0)M^\theta(\Ga)\dg] \la{P_Gamma}
}
where
\bea{
M^\theta(\Ga) \aeqd W^\theta(\tau, t_\mN)
\Big(\prod_{\al=1}^\mN L_{k_\al}^\theta(t_\al) W^\theta(t_\al,t_{\al-1}) \Big)
}
with $t_0=0$.
$W^\theta(t,s)$ is defined by 
\bea{
\f{\p W^\theta(t,s)}{\p t} \aeq \Big(-iH_S^\theta -\half \sum_k (L_k^\theta)\dg L_k^\theta \Big)W^\theta(t,s)
}
under $W^\theta(s,s)=1$ with 
\bea{
H_S^\theta \aeqd (1+\theta)H_S \com
L_k^\theta \defe \sqrt{1+\theta}L_k. \la{def_theta}
}
Here, $\theta$ is the time-rescaling real parameter. 
The Fisher information $F$ is defined by 
\bea{
F \aeqd -\bra \p_\theta^2 \ln \mP^\theta(\Ga) \big \vert_{\theta=0} \ket ,\la{def_F_our}
}
where $\bra \bu \ket $ denotes the expected value for the probability distribution $\mP^\theta(\Ga)\vert_{\theta=0}$. 
$F$ is bounded as (see Appendix \ref{A_SLD})
\bea{
F \le \mI. \la{I_J_relation}
}
$\mI$ is given by \re{def_FI}.
The two-sided GKSL equation governing $\rho^{\theta_1,\theta_2}$ is given by
\bea{
\f{d\rho^{\theta_1,\theta_2}(t)}{dt}\aeq \mL^{\theta_1,\theta_2}(t) \rho^{\theta_1,\theta_2}(t) \la{two_2}
}
under $\rho^{\theta_1,\theta_2}(0)=\rho(0)$ \cite{SLD}. 
Here, 
\bea{
&\mL^{\theta_1,\theta_2}(t) \bu \defe -iH_S^{\theta_1} \bu +\bu i H_S^{\theta_2} 
+\sum_k \Big\{ L_k^{\theta_1} \bu (L_k^{\theta_2})\dg \no\\
&\hs{15mm}-\half \Big[ (L_k^{\theta_1})\dg L_k^{\theta_1}\bu + \bu  (L_k^{\theta_2})\dg L_k^{\theta_2} \Big] \Big\}. \la{two_2L}
}

$F$ can be calculated only numerically.
First, we discretize time and introduce
\bea{
\Om_0^\theta \aeqd 1+\Big(-i H_S^\theta -\half  \sum_{m=1}^M  (L_m^\theta)\dg L_m^\theta \Big)\Dl t  , \la{Om_0} \\
\Om_m^\theta \aeqd L_m^\theta \sqrt{\Dl t}  \ \ (m=1,\cdots, M).\la{Om_mu}
}
Here, $\Dl t\defe \tau/N$ with a sufficiently large integer number $N$. 
$M$ is the number of the jump operator of the GKSL equation. 
The probability of quantum trajectory $\mP^\theta(\Ga)$ is
\bea{
&\hs{-3mm}P^\theta(\{m_i \}) \no\\
\aeqd \tr_S[\Om_{m_{N-1}}^\theta \cdots \Om_{m_0}^\theta \rho(0)(\Om_{m_0}^\theta)\dg \cdots (\Om_{m_{N-1}}^\theta)\dg ] \la{pro_Nu}
}
in $N \to \infty$.
Then by quantum jump method \cite{QJ}, we numerically construct each trajectory and calculate
\bea{
\tl F \defe -\sum_{m_0, \cdots,m_{N-1}=0}^M P^\theta(\{m_i \}) \p_\theta^2 \ln P^\theta(\{m_i \}) \Big\vert_{\theta=0} \la{F2}
}
\rd{which is the discrete version of $F$. 
In $N \to \infty$ limit, $\tl F$ becomes $F$}.  

Note that \re{P_Gamma} is slightly different from the original definition in Ref.\cite{Vu22}, in which the initial state is decomposed as  
\bea{
\rho(0) = \sum_\al p_\al \ke{\al}\br{\al} \la{deco}
}
where $\{\ke{\al}\}$ are normalized but need not be orthogonalized.
The probability of quantum trajectory is defined by
\bea{
&\hs{-3mm}P^\theta(\al, \{m_i \}) \no\\
\aeqd p_\al \tr_S[\Om_{m_{N-1}}^\theta \cdots \Om_{m_0}^\theta 
\ke{\al}\br{\al}(\Om_{m_0}^\theta)\dg \cdots (\Om_{m_{N-1}}^\theta)\dg ]. \la{18}
}
The associated Fisher information is 
\bea{
\tl F^\pr 
\aeqd-\sum_\al  \sum_{m_0, \cdots,m_{N-1}=0}^M P^\theta(\al, \{m_i \}) \no\\
&\spa \times \p_\theta^2 \ln P^\theta(\al, \{m_i \}) \Big\vert_{\theta=0}. \la{def_tl_F^pr}
}
Vu-Saito's original Fisher information $F^\pr$ is $N \to \infty$ limit of $\tl F^\pr$. 
In general, $F^\pr$ does not coincide with $F$ and 
depends on the decomposition \re{deco}.
However the upper bound derived from the quantum Cram\'{e}r-Rao theorem \cite{94, FN}
 is independent of the definitions and $F^\pr \le \mI$ holds (Appendix \ref{A_SLD}).

\section{SLD Fisher information} \la{s_Fisher}

\subsection{Long time approximation}

In \rd{Refs.\cite{QTUR3} and} \cite{Vu22}, the SLD Fisher information $\mI$ is calculated in the limit of the long-time
when $H_S$ and $L_k$ are time-independent. 
\rd{The SLD Fisher information $\mI$ can be rewritten as 
\bea{
\mI \aeq 4\p_{\theta_1}\p_{\theta_2}\ln \tr_S \rho^{\theta_1,\theta_2}(\tau)\bv{\theta_1=0=\theta_2} .
}
If $H_S$ and $L_k$ are time-independent, $\mI$ becomes
\bea{
\mI\aeq 4\tau \p_{\theta_1}\p_{\theta_2}\lm(\theta_1, \theta_2)\bv{\theta_1=0=\theta_2}+O(1) \la{mI_in_long}
}
in the limit of the long-time.
Here, $\lm(\theta_1, \theta_2)$ is the eigenvalue of $\mL^{\theta_1,\theta_2}$ which satisfies $\lm(0,0)=0$. 
Based on this, Refs.\cite{QTUR3} and \cite{Vu22} have derived 
\bea{
\mI \aeqap \tau(\dot{B}_\RM{ss}+\dot{Q}_+) ,\\
\dot{B}_\RM{ss} \aeqd \sum_k \tr_S[L_k \dg L_k \rho^\RM{ss}]  ,\\
\dot{Q}_+ \aeqd -4\Big( \tr_S\Big[\mL_2\mR \mL_1 \rho^\RM{ss}\Big]
 +\tr_S\Big[\mL_1\mR \mL_2\rho^\RM{ss}\Big] \Big) \la{dot_Q_+_long}
} 
with
\bea{
\mL_1 \bu \aeqd -iH_S\bu +\half\sum_k \Big[ L_k\bu L_k \dg - L_k\dg L_k \bu \Big],\\
\mL_2 \bu \aeqd \bu iH_S+ \half \sum_k \Big[ L_k \bu  L_k \dg -\bu L_k\dg L_k\Big] .
}
Here, $\rho^\RM{ss}$ is the steady state and $\mR$ is the pseudo-inverse of the Liouvillian (see \re{def_R_0})}.

\subsection{Our result}

Here, \rd{when $H_S$ and $L_k$ dependent on time,} we derive an expression of $\mI$ for arbitrary times \rd{based on \re{def_FI}}.
The first term of \re{def_FI} is reduced to the sum of the dynamical activity $B(\tau)$ and two double integrals (Appendix \ref{A_B}):
\bea{
\p_{\theta_1}\p_{\theta_2}\mC(\theta_1,\theta_2) \bv{\theta_1=0=\theta_2} \aeq \f{1}{4}B(\tau)+I_1+I_2 ,\la{def_I}
}
where
\bea{
B(t) \aeqd \int_0^t ds \ \sum_k \tr_S [L_k(s)\rho(s)L_k(s)\dg] ,\\
I_1 \aeqd  \int_0^\tau ds \int_0^s du \  \tr_S\Big[\mL_{2}(s) \mU(s,u) \mL_{1}(u) \rho(u)\Big],\\
I_2 \aeqd  \int_0^\tau ds \int_0^s du \  \tr_S\Big[\mL_{1}(s) \mU(s,u) \mL_{2}(u) \rho(u)\Big].
}
Here, $\mU(s,u)$ is defined by 
\bea{
\f{\p \mU(s,u)}{\p s} \aeq \mL(s)\mU(s,u) 
}
with $\mU(u,u)=1$. 
The second term of \re{def_FI} is calculated as
\bea{
&\hs{-3mm}-\p_{\theta_1}\mC(\theta_1,\theta_2)\p_{\theta_2}\mC(\theta_1,\theta_2) \bv{\theta_1=0=\theta_2} \no\\
\aeq -\prod_{i=1}^2\int_0^\tau ds \ \tr_S\Big[\mL_{i}(s)  \rho(s)\Big] \no\\
\aeq -\Big(\int_0^\tau ds \ \tr_S\Big[H_S(s)\rho(s)\Big] \Big)^2
=:I_3. \la{def_I_3}
}
Here, we used 
\bea{
\tr_S[\mL_1\bu] \aeq -i\tr_S[H_S\bu]  =-\tr_S[\mL_2\bu] .\la{dbr_1_mL_i}
}
Thus, the SLD Fisher information is given by
\bea{
\mI \aeq B(\tau)+Q_+(\tau) ,\\
Q_+ \aeqd 4(I_1+I_2+I_3).
}
Here, $Q_+$ is the upper bound of the quantum correction $\mQ \defe F-B$. 

$Q_+(t)$ can be written as (Appendix \ref{A_B})
\bea{
Q_+=2\RM{Re}(Q_a) \la{Q_+_re}
}
with
\bea{
&\hs{-3mm}Q_a(t) \no\\
&\hs{-3mm}\defe 4\int_0^t ds \int_0^s du \ \tr_S\Big[\mL_2(t) \mU(s,u) \mP(u)\mL_1(u)\rho(u)\Big].
}
Here, $\RM{Re}(x)$ is the real part of $x$ and 
\bea{
\mP(u)\bu \defe \bu-\rho(u)\tr_S[\bu] .
}
The double time integral is given by the following coupled differential equations,
\bea{
\f{dQ_a}{dt} \aeq 4\tr_S[\mL_2(t) q_a(t)] , \la{main_1} \\
\f{dq_a}{dt} \aeq \mP(t)\mL_1(t)\rho(t)+\mL(t) q_a(t) \la{main_2}
} 
with $Q_a(0)=0$, $q_a(0)=0$, and the GKSL equation \re{QME}.

\re{main_1}, \re{main_2}, and \re{QME} can be solved by standard numerical methods such as the Runge-Kutta method.
\re{Q_+_re}, \re{main_1}, and \re{main_2} are the first main results of this paper. 
In Appendix \ref{A_correction}, we check $Q_+$ in the long-time region reproduces \rd{\re{dot_Q_+_long}}. 

The lower bound of $\mQ$ is also given by a double time integral (Appendix \ref{s_Lower}):
\bea{
\mQ \ge Q_- \aeqd B(\tau)^2 \no\\
&\spa-2\int_0^\tau ds\int_0^s du \ \tr_S[\hat{\ga}(s) \mU(s,u)\hat{\ga}(u) \rho(u)] \no\\
&=: B(\tau)^2 +Q_c(\tau). \la{Q_-}
}
Here, 
\bea{
\hat{\ga}(t) \bu \aeqd \Big(-i H_S -\half  \sum_k  L_k\dg L_k\Big) \bu \no\\
&\spa+  \bu \Big(i H_S -\half  \sum_k  L_k\dg L_k\Big). \la{def_ga}
}
$Q_c(t)$ can be calculated in the same way as $Q_a$.

\section{Numerical analysis} \la{s_Numerical}

In the following, we suppose that $H_S$ and $L_k$ are time-independent.

\begin{figure*}[t]
\begin{center}
  \begin{minipage}[b]{0.45\linewidth}
    \centering
    \includegraphics[keepaspectratio, width = 1 \columnwidth]{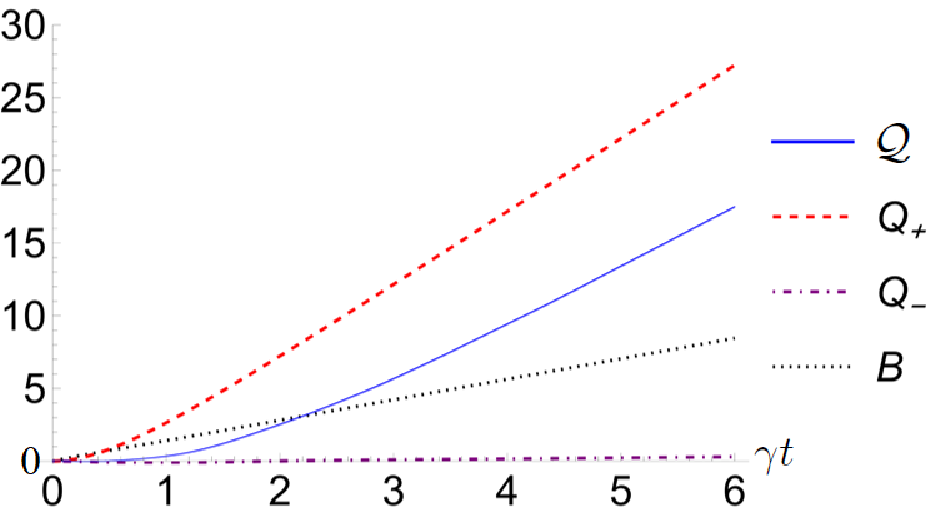}
    \subcaption{$\ga t \le 6$}
  \end{minipage}
  \begin{minipage}[b]{0.45\linewidth}
    \centering
    \includegraphics[keepaspectratio, width = 1 \columnwidth]{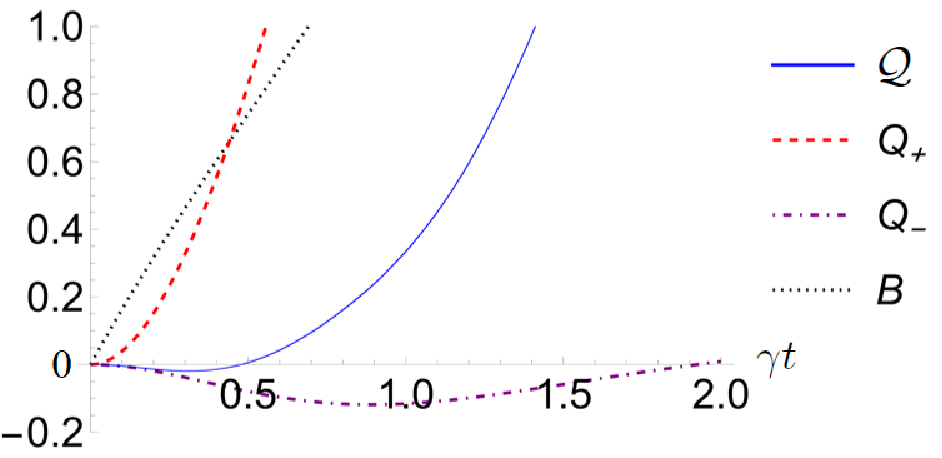}
    \subcaption{$\ga t \le 2$}
  \end{minipage}
  \caption{\label{Fig_Q}$\mQ$, $Q_+$, $Q_-$, and $B$ for (a) $\ga t \le 6$ and (b) $\ga t \le 2$.
  The horizontal axes are $\ga t$. 
We set $\Om=\ga$, $\Dl=0.25\ga$, $n=1$ and 
$\rho(0)=\half(1+0.2\sig_x+0.3\sig_y-0.4\sig_z)$. 
Here, $\sig_i$ is the Pauli matrix ($\sig_y=-i\ke{0}\br{1}+i\ke{1}\br{0}$, etc.).
}
\end{center}
\end{figure*}

As an example, we consider a system described by
\bea{
H_S \aeq \Dl \ke{1}\br{1}+\Om(\ke{0}\br{1}+\ke{1}\br{0}) ,\la{Nu1}\\
L_1 \aeq \sqrt{\ga n}\ke{1}\br{0} ,\la{Nu2}\\
L_2 \aeq \sqrt{\ga (n+1)}\ke{0}\br{1}.\la{Nu3}
}
Figure \ref{Fig_Q} shows the time dependence of the dynamical activity $B$ and the quantum correction $\mQ$
and its upper  and lower bounds ($Q_+$ and $Q_-$). 
$F$ was calculated by using \re{Om_0}, \re{Om_mu}, \re{pro_Nu} and \re{F2}
with the quantum jump method \cite{QJ}. 
In Fig.\ref{Fig_Q}, we set $\ga \Dl t = 0.001$ and used $10^6$ trajectories. 
In both panels, we observe $Q_-\le \mQ \le Q_+$ holds.
Figure \ref{Fig_Q} (a) shows that the quantum correction $\mQ$ of this example is comparable to the dynamical activity $B$. 
At short times, $\mQ$ also takes negative values and $Q_-$ provides a good lower bound as Fig.\ref{Fig_Q} (b) shows. 
After the state relaxes to the steady state, \rd{$Q_\pm$} increases with a slope of \rd{$\dot{Q}_\pm$}.
Here, 
$\rd{\dot{Q}_\pm}\defe \lim_{\tau \to \infty} \rd{Q_\pm}/\tau$ \rd{are} given by \cite{Vu22} (Appendix \ref{A_D})
\begin{widetext}
\bea{
\dot{Q}_+ \aeq \f{8\fA}{\ga y^3[4(\Dl^2+2\Om^2)+\ga^2y^2]^3} ,\la{dot_Q}\\
\rd{\dot{Q}_-} \aeq \rd{\f{2\ga \fB}{y^3[4(\Dl^2+2\Om^2)+\ga^2y^2]^3} } ,\la{dot_Q_-} \\
\fA \aeqd \Dl^2x(4\Dl^2+\ga^2y^2)^3 
+ 8\Om^2x(4\Dl^2+\ga^2y^2)^2(6\Dl^2+\ga^2y^2) \no\\
&\spa + 16\Om^4\Big(\ga^2\Dl^2y^2(100x+1)+\ga^4y^4(12x+1)+4\Dl^4(52x+1) \Big) \no\\
&\spa+256\Om^6\Big(\ga^2(6x+1)y^2+2\Dl^2(12x+1)\Big) +1024\Om^8y^2 ,\\
\rd{\fB} \aeqd \rd{-x(4\Dl^2+\ga^2y^2)^3+16\Om^2x(-16\Dl^4+\ga^4y^4)+16\Om^4 y^2(-4\Dl^2+3\ga^2y^2)}
}
\end{widetext}
with $x\defe n(n+1)$ and $y\defe 2n+1$.  
\rd{$\dot{Q}_+$ is non-negative. 
In general, $Q_\pm$ saturates when $\dot{Q}_\pm = 0$. }

\section{Quantum speed limit} \la{s_speed}

In this section, we discuss a quantum speed limit derived by Hasegawa \cite{H22}:
\bea{
\half \int_{t_1}^{t_2} dt \ \sqrt{\mJ(t)} \ge D(\sig(t_1), \sig(t_2)) .\la{MTR}
}
Here, $\sig(t) \defe \ke{\Psi_\tau(\tau; t)}\br{\Psi_\tau(\tau; t)}$. 
\re{MTR} is the Mandelstam-Tamm relation \cite{Mandelstam, review_MT} applied to a state $\ke{\Psi_\tau(s;t)}$ defined by 
\bea{
\ke{\Psi_\tau(s;t)} \defe V_{\tau}(s;t) \ke{\tl \psi(0)} \otimes \ke{0}
}
with 
\bea{
&\hs{-3mm}V_{\tau}(s;t) \defe\RM{T} \exp \Big[\int_0^s du \ \Big\{-i\f{t}{\tau} H_S\Big(\f{t}{\tau}u\Big) \no\\
&\hs{5mm}+ \sqrt{\f{t}{\tau}} \sum_k [L_k\Big(\f{t}{\tau}u\Big)\otimes \phi_k\dg (u) \no\\
&\hs{17mm}-L_k\Big(\f{t}{\tau}u\Big)\dg \otimes \phi_k (u)] \Big\} \Big]. \la{def_V_op}
}
Here, $\ke{\tl \psi(0)}$ is a purification of $\rho(0)$ ($\tr_A[\ke{\tl \psi(0)}\br{\tl \psi(0)}]=\rho(0)$ where $A$ is the ancilla system).
$\RM{T}$  is the time ordering operator.
$\{\phi_k(t)\}$ are field operators having the canonical commutation relation
\bea{
[\phi_k(t), \phi_l\dg (s)] \aeq \dl_{kl}\dl(t-s)
}
and $\ke{0}$ is the vacuum state for the fields. 
$\mJ(t)$ is the SLD Fisher information for time (see \re{SLD_pre}):
\bea{
\mJ(t) \aeqd  4[\bra \p_t \Psi_\tau(\tau;t) \ke{\p_t \Psi_\tau(\tau;t)} \no\\
&-\bra \p_t \Psi_\tau(\tau;t) \ke{\Psi_\tau(\tau;t)}\bra \Psi_\tau(\tau;t) \ke{\p_t \Psi_\tau(\tau;t)}] \la{def_mJ}
}
where $\ke{\p_t \Psi_\tau(\tau;t)} \defe \p_t \ke{\Psi_\tau(\tau;t)}$. 
In Mandelstam-Tamm relation, $\mJ(t)/4$ is reduced to the energy fluctuation. 
In Hasegawa's theory, the Hamiltonian becomes $i\f{\p V_\tau(\tau;t)}{\p t}V_\tau(\tau;t)\dg$, 
which does not related to the real system dynamics and thus the physical meaning may not be so clear. 
$D(\rho, \sig)$ is the Bures angle:
\bea{
D(\rho, \sig) \aeqd \cos^{-1}F(\rho, \sig) ,\la{def_Bures}\\
F(\rho, \sig) \aeqd \tr\sqrt{\sqrt{\rho}\sig \sqrt{\rho}}=F(\sig, \rho).
}
 $F(\rho, \sig) $ is the fidelity \cite{NC}. 
Because of the contractivity of the Bures angle (p.414 of Ref.\cite{NC}) and $\rho(t)=\tr_{AB} \sig(t)$ ($B$ is the field system), 
 \bea{
 D(\sig(t_1), \sig(t_2)) \ge D(\rho(t_1), \rho(t_2)) \la{MTR2}
 }
holds. 
Note that the trace distance $T(\rho_1,\rho_2) \defe \half \tr_S \sqrt{(\rho_1-\rho_2)^2}$ 
is smaller than the Bures angle $D(\rho_1,\rho_2) \ge T(\rho_1,\rho_2)$ \cite{NC}. 
For  time-independent $H_S$ and $L_k$, we recognize that
the two SLD Fisher information \re{def_FI} and \re{def_mJ} are connected  (Appendix \ref{A_Hasegawa}) 
\bea{
\mJ(t)=\f{\mI(t)}{t^2}=\f{B(t)+Q_+(t)}{t^2}. \la{MTR3}
}

The quantum correction $Q_+$ can be eliminated in the interaction picture $\tl \rho(t)\defe e^{iH_S t}\rho(t)e^{-iH_S t}$
when 
\bea{
[L_k, H_S] \aeq \om_k L_k, 
}
where $\om_k$ is a real number. 
The quantum master equation for $\tl \rho(t)$ is given by
\bea{
\f{d\tl \rho}{dt} \aeq \sum_k \Big[L_k \tl \rho(t) L_k \dg - \half \{  L_k \dg L_k ,\tl \rho(t) \} \Big].
}
Repeating the arguments from \re{MTR} to \re{MTR3}, we obtain a quantum speed limit of the system expressed with the dynamical activity:
\bea{
l(t_1,t_2) \defe \half \int_{t_1}^{t_2} dt \ \f{\sqrt{B(t)}}{t} \ge D(\tl \rho(t_1), \tl \rho(t_2)) . \la{main_3}
} 
Here, $0 \le t_1 \le t_2$. 
\re{main_3} is the second main result of this paper.

As an instance, we consider a spinless quantum dot coupled to a single lead, 
\bea{
\f{d\rho}{dt} = -i[H_S, \rho] +\ga[1-f(\ep)]\mD[a]( \rho)
+ \ga f(\ep)\mD[a\dg](\rho) \la{QME_QD}
}
where $H_S=\ep a\dg a$ and $\mD[X](\bu) \defe X\bu X\dg-\half \{X\dg X, \bu \}$. 
Here, $a$ is the annihilation operator of the electron of the system, 
$\ep$ is the energy level of the system, 
$f(\ep) = \f{1}{e^{\be\ep}+1}$ is the Fermi distribution, 
$\be$ is the inverse temperature of the lead, and  
$\ga$ is the coupling strength. 
The jump operators are $L_1=\sqrt{\ga[1-f(\ep)]}a$ and $L_2=\sqrt{\ga f(\ep)} a\dg$ with 
$\om_1=\ep$ and $\om_2=-\ep$. 
(See Appendix \ref{A_QD} for detailed calculations).
Figure \ref{Fig_1}(a) shows the Bures  angle $D$, the trace distance $T$, and geometric length $l$ as functions of final time $t_2$. 
In Fig.\ref{Fig_1}, we set the initial time $\ga t_1=2$. 
Figure \ref{Fig_1}(b) shows that the bound achievement ratio $D(\tl \rho(t_1), \tl \rho(t_2))/l(t_1.t_2)$ becomes greater than 0.8 around $\ga t_2=2$.  
In Fig.\ref{Fig_2}, we showed the results when the initial time $\ga t_1=0$. 
Around $\ga t=0$, because of $B(t) \propto t$ and $l(0,t) \propto \sqrt{t}$, 
the bound achievement ratio grows in square root of time  $D(\tl \rho(0), \tl \rho(t))/l(0,t)  \propto \sqrt{t}$. 
The square root dependence makes the upper bound looser in the short time regime as shown in Fig.\ref{Fig_2}(a).


\begin{figure*}[t]
\begin{center}
  \begin{minipage}[b]{0.45\linewidth}
    \centering
    \includegraphics[keepaspectratio, width = 1 \columnwidth]{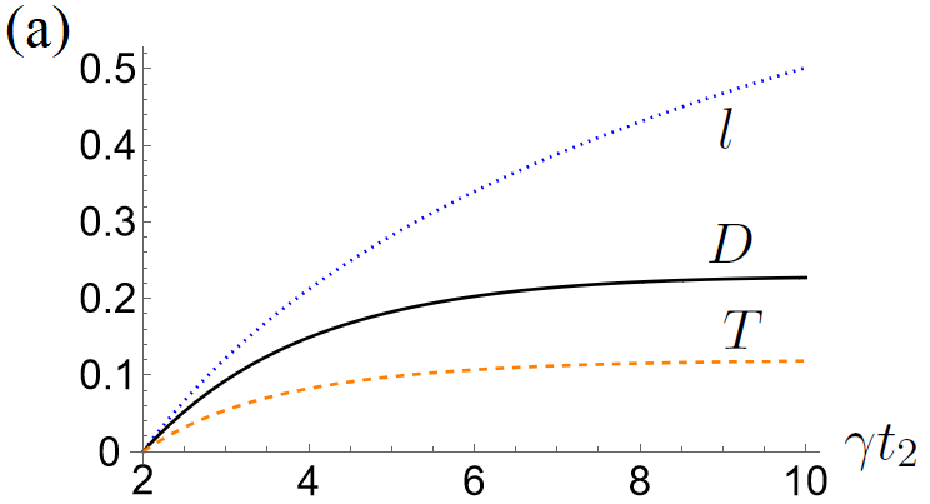}
    
  \end{minipage}
  \begin{minipage}[b]{0.45\linewidth}
    \centering
    \includegraphics[keepaspectratio, width = 1 \columnwidth]{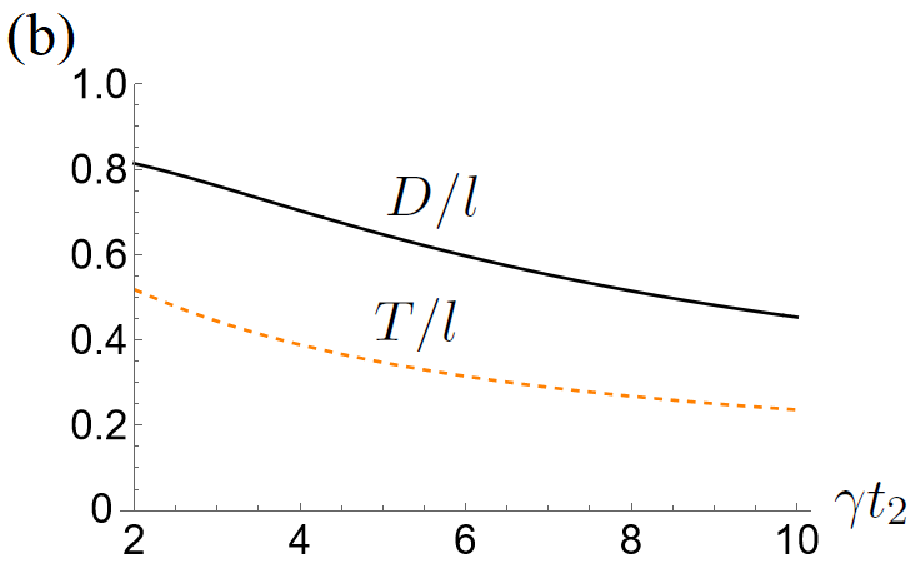}
    
  \end{minipage}
  \caption{\label{Fig_1}(a) $l(t_1.t_2)$, $D(\tl \rho(t_1), \tl \rho(t_2))$, and $T(\tl \rho(t_1), \tl \rho(t_2))$, 
  (b) $D(\tl \rho(t_1), \tl \rho(t_2))/l(t_1.t_2)$, and $T(\tl \rho(t_1), \tl \rho(t_2))/l(t_1.t_2)$ for $\ga t_1=2$. 
  The horizontal axes are $\ga t_2$. 
  $\be \ep =10$, $\rho(0)=\half(1-0.5\sig_x+0.3\sig_y+0.2\sig_z)$.  $\sig_i$ is the Pauli matrix.
}
\end{center}
\end{figure*}

\begin{figure*}[t]
\begin{center}
  \begin{minipage}[b]{0.45\linewidth}
    \centering
    \includegraphics[keepaspectratio, width = 1 \columnwidth]{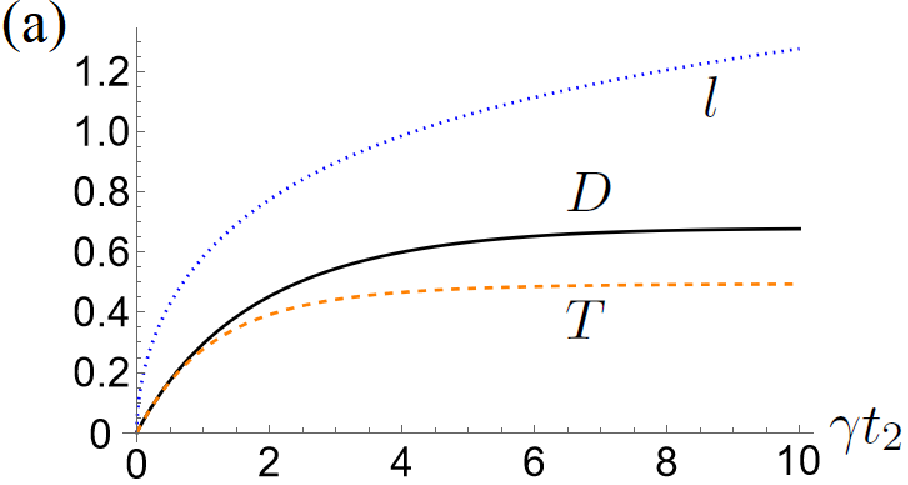}
    
  \end{minipage}
  \begin{minipage}[b]{0.45\linewidth}
    \centering
    \includegraphics[keepaspectratio, width = 1 \columnwidth]{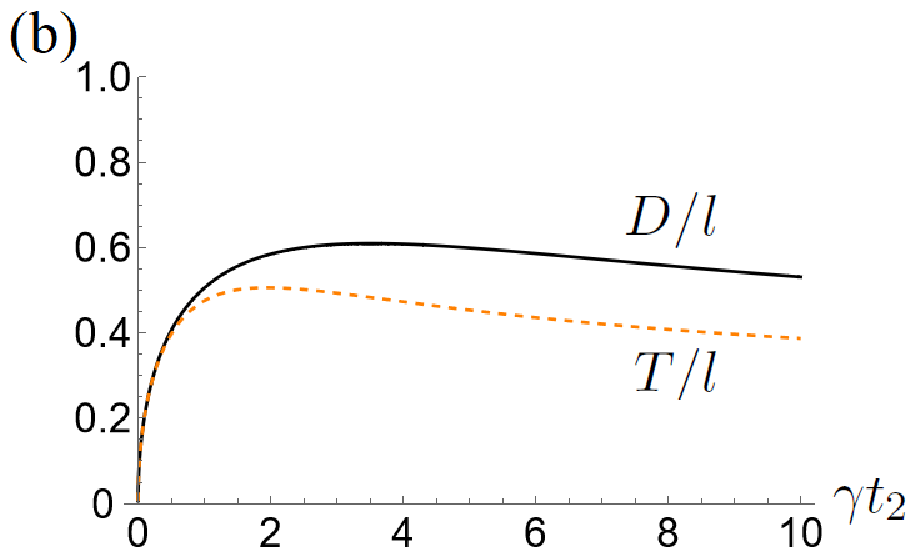}
    
  \end{minipage}
  \caption{\label{Fig_2}(a) $l(t_1.t_2)$, $D(\tl \rho(t_1), \tl \rho(t_2))$, and $T(\tl \rho(t_1), \tl \rho(t_2))$, 
  (b) $D(\tl \rho(t_1), \tl \rho(t_2))/l(t_1.t_2)$, and $T(\tl \rho(t_1), \tl \rho(t_2))/l(t_1.t_2)$ for $\ga t_1=0$. 
  The horizontal axes are $\ga t_2$. 
  Other parameters are same with Fig.\ref{Fig_1}.
}
\end{center}
\end{figure*}


\rd{Naive extensions of \re{SFS} and \re{SL_A} to the quantum regime may be,} 
\bea{\sqrt{\f{B(\tau)\sig}{2}}\aeqge T(\tl \rho(0), \tl \rho(\tau)) ,\\
B(\tau) \aeqge T(\tl \rho(0), \tl \rho(\tau))  .
}
However, we numerically checked that they fail even in the spinless quantum dot due to the quantum effect.
Actually the quantum extensions of \re{SFS} contain nontrivial quantum corrections \cite{Funo19, Vu23, Nakajima22}. 
However\rd{,} the quantum extension of \re{SL_A} has been less investigated. 
We speculate 
\bea{
\sqrt{B(\tau)} \aeqge D(\tl \rho(0), \tl \rho(\tau)) \la{66}
}
from the discussion of Fig. \ref{Fig_2}. 
However, from \re{main_3}, we only could derive a looser bound, 
\bea{
\sqrt{b_\RM{max}(\tau)\tau} \ge D(\tl \rho(0), \tl \rho(\tau))  \la{b_max}
}
with 
\bea{
b_\RM{max}(\tau) \defe \max_{0\le t \le \tau} \f{1}{t}B(t).
} 
Here, we used $\sqrt{b_\RM{max}(\tau)\tau}  \ge l(0, \tau)$. 
Thus \re{66} would be correct in the short time limit. 
Derivation of a simple quantum speed limit at zero initial time is a future work. 
The extension to the first passage time \cite{FPT1, FPT2, FPT3} for open quantum systems
is also an interesting problem.

\section{Summary} \la{s_summary}

We investigated the symmetric logarithm derivative (SLD) Fisher information,
which appears in the context of KUR and is the upper bound of the Fisher information of the quantum trajectory for the time-rescaling parameter. 
For a finite time and arbitrary initial state, we derived a concise expression of the SLD Fishder information using a double time integral, 
which can be calculated by numerically solving coupled first-order ordinary differential equations.  
We also derived a simple lower bound of the Fisher information for the probability of quantum trajectory.
Furthermore, we pointed out that for time-independent system, the SLD Fisher information divided by time squared is identical to the SLD Fisher information 
appeared in the Mandelstam-Tamm speed limit by Hasegawa \cite{H22}.
Based on this observation, we showed that when the jump operators connects energy eigenstates, 
the upper bound of the Bures angle between the initial and final states in the interaction picture 
is expressed with the square root of the dynamical activity. 

\acknowledgments

We acknowledge helpful discussions with Tan Van Vu. 
This work was supported by JSPSKAKENHI Grants No. 18KK0385, and No. 20H01827. 

\appendix

\section{SLD Fisher information and the two-sided GKSL equation} \la{A_SLD}

\subsection{SLD Fisher information}

We consider $n$ real parameters $\theta=(\theta^1, \cdots, \theta^n)$ and a state $\rho^\theta$. 
The SLD $S_i^\theta$ is  defined by 
\bea{
\p_i \rho^\theta \aeq \half (\rho^\theta S_i^\theta + S_i^\theta \rho^\theta) \com S_i^\theta=(S_i^\theta)\dg .
}
Here, $\p_i =\p/\p\theta^i$. 
Although $S_i^\theta$ is not unique in general, the SLD Fisher information matrix
\bea{
J_{ij}^\theta \defe \half \tr \Big[\rho^\theta \{ S_i^\theta, S_j^\theta \} \Big] 
}
is unique \cite{FN}. 
$J_{ij}^\theta$ can be rewritten as $J_{ij}^\theta=\tr[\p_i \rho^\theta S_j^\theta]$. 

In the following, we consider a pure state $\rho^\theta$. 
Differentiating $(\rho^\theta)^2=\rho^\theta$, we obtain
\bea{
\p_i \rho^\theta \aeq (\p_i \rho^\theta)\rho^\theta + \rho^\theta\p_i \rho^\theta.
}
Thus,  $2 \p_i \rho^\theta$ is an SLD. 
Using this relation and denoting $\rho^\theta=\ke{\psi^\theta}\br{\psi^\theta}$, we obtain
\bea{
J_{ij}^\theta = 4  \RM{Re}  \Big[\bra \p_i \psi^\theta \ke{\p_j \psi^\theta}
- \bra \p_i \psi^\theta \ke{\psi^\theta}\bra \psi^\theta \ke{\p_j \psi^\theta}  \Big].
}
Here,  $\ke{\p_i \psi^\theta}\defe \p_i \ke{\psi^\theta}$. 
For $n=1$, $J^\theta \defe J_{11}^\theta $ becomes
\bea{
J^\theta =4  \big[\bra \p_\theta \psi^\theta \ke{\p_\theta \psi^\theta}
- \bra \p_\theta \psi^\theta \ke{\psi^\theta}\bra \psi^\theta \ke{\p_\theta \psi^\theta}  \big]. \la{SLD_pre}
}

\subsection{Continuous measurement}

We introduce $(M+1)$-dimensional Hilbert space $\mH$ with an orthonormal basis $\{\ke{m}\}_{m=0}^M$ and 
a fictitious environment system $E$ of which Hilbert space is $\mH^{\otimes N}$. 
We consider a combined system of $S$, the ancilla system $A$, and $E$. 
We suppose that the initial state of the combined system is $\ke{\tl \psi(0)}\otimes \ke{0_{N-1}, \cdots, 0_1, 0_0}$. 
Here, $\ke{\tl \psi(0)}$ is the purification of $\rho(0)$ (\textit{i.e.} $\tr_A[\ke{\tl \psi(0)}\br{\tl \psi(0)}]=\rho(0)$), and
$\ke{0_{N-1}, \cdots, 0_1, 0_0} = \otimes_{i=0}^{N-1} \ke{0}_i$. 
For each $i=0,1,\cdots, N-1$, an environmental subspace $i$ interacts with system $S$ during the time interval $[i\Dl t, (i + 1)\Dl t]$ via a unitary operator $U_i$. 
Here, $\Dl t \defe \tau/N$. 
The state of the combined system at time $\tau$ is given by
\bea{
\ke{\psi^\theta} \aeq U_{N-1}\cdots U_{1}U_0 \ke{\tl \psi(0)}\otimes \ke{0_{N-1}, \cdots, 0_1, 0_0} \no\\
\aeq \sum_{m_0, \cdots,m_{N-1}=0}^M \Om_{m_{N-1}}^\theta \cdots \Om_{m_0}^\theta
\ke{\tl \psi(0)} \no\\
&\hs{27mm} \otimes \ke{m_{N-1}, \cdots, m_1, m_0}
}
where $\Om_{m_i}^\theta$ is defined by
\bea{
\br{k_S}\Om_{m_i}^\theta\ke{k_S^\pr} \aeq \br{k_S}{}_i \br{m_i} U_{i}\ke{k_S^\pr}\ke{0}_i .
}
$\ke{k_S}$ and $\ke{k_S^\pr}$ are bases of the system.
We suppose that $\Om_m^\theta$ are the same as \re{Om_0} and \re{Om_mu}.
The Fisher information associated with POVM (positive operator valued measure) \cite{NC} $\mM$ is denoted by $I(\theta, \mM)$. 
If we put $\mM_0(\{m_i \}) \defe 1_{SA} \otimes \ke{m_{N-1}, \cdots, m_1, m_0} \br{m_{N-1}, \cdots, m_1, m_0}$ 
($1_{SA}$ is the identity operator of $SA$), 
the outcome is given by
\bea{
\tr [\mM_0(\{m_i \}) \ke{\psi^\theta}\br{\psi^\theta}] =P^\theta(\{m_i \}) .
}
Thus, \rd{$\tl F$ defined by \re{F2} is given by}
\bea{
\tl F \aeq I(0, \mM_0(\{m_i \})) .\la{F_1a}
}
Because of the quantum Cram\'{e}r-Rao theorem \cite{94, FN}, 
\bea{
I(\theta, \mM) \le J^\theta \la{F_2}
}
holds. 
Here, $J^\theta$ is the SLD Fisher information given by \re{SLD_pre}. 
Using \re{F_1a}, \re{F_2} and \re{SLD_pre}, we obtain 
\bea{
F \aeqle \mI = 4  \Big[\f{\p^2}{\p \theta_1 \p \theta_2} \bra \psi^{\theta_2} \ke{ \psi^{\theta_1}} \no\\
&\spa - \Big(\f{\p}{ \p \theta_2} \bra \psi^{\theta_2}  \ke{\psi^{\theta_1}} \Big) 
\Big(\f{\p}{ \p \theta_1}\bra \psi^{\theta_2} \ke{\psi^{\theta_1}} \Big) \Big]
\Big \vert_{\theta_1=0=\theta_2} .
}
Here, 
\bea{
\bra \psi^{\theta_2} \ke{ \psi^{\theta_1}} \aeq \tr_S \rho^{\theta_1,\theta_2}(\tau) ,\\
\rho^{\theta_1,\theta_2}(\tau) \aeqd \tr_{AE}[\ke{\psi^{\theta_1}}\br{\psi^{\theta_2}}].
}
The time evolution equation of $\rho^{\theta_1,\theta_2}(t)$ is given by \re{two_2} \cite{SLD}.
Then, we obtain \re{I_J_relation}.

To obtain \re{18}, we adopt a purification
\bea{
\ke{\tl \psi(0)} = \sum_\al \sqrt{p_\al} \ke{\al} \otimes \ke{\varphi_\al}_A
}
with ${}_A\bra\varphi_\al \ke{\varphi_\be}_A=\dl_{\al\be}$ and the POVM 
$\mM_0(\al, \{m_i \}) \defe \ke{\varphi_\al}_A{}_A\br{\varphi_\al} \otimes \ke{m_{N-1}, \cdots, m_1, m_0} \br{m_{N-1}, \cdots, m_1, m_0}$. 
Then the outcomes becomes \re{18}:
\bea{
 \tr [\mM_0(\al, \{m_i \}) \ke{\psi^\theta}\br{\psi^\theta}] =P^\theta(\al, \{m_i \}) .
}
The associated Fisher information \rd{defined by \re{def_tl_F^pr}} is given by
\bea{
\tl F^\pr \aeq I(0, \mM_0(\al, \{m_i \})).
}
Then, $F^\pr \le \mI $ holds.

\section{Derivation of \re{def_I}} \la{A_B}

In the following, we use the Liouville space.
An arbitrary linear operator $X$ of the system is described by a vector $\dke{X}$. 
The inner product is defined by $\dbra Y \dke{X}\defe \tr_S(Y\dg X)$.
In particular, $\dbra 1 \dke{X}=\tr_S(X)$. 
An arbitrary linear super-operator of  the system is described by an operator of Liouville space. 
The conservation of the probability leads to $\dbr{1}\mL(t)=0$.

$\mC(\theta_1,\theta_2)$ is given by 
\bea{
\mC(\theta_1,\theta_2) \aeq \dbr{1}\mU^{\theta_1,\theta_2}(\tau,0)\dke{\rho(0)}
}
where $\mU^{\theta_1,\theta_2}(u,s)$ is defined by 
\bea{
\f{\p \mU^{\theta_1,\theta_2}(u,s)}{\p u} \aeq \mL^{\theta_1,\theta_2}(u)\mU^{\theta_1,\theta_2}(u,s)
}
with $\mU^{\theta_1,\theta_2}(s,s)=1$. 
Note that $\mU^{0,0}(u,s)=\mU(u,s)$. 
The first derivative \rd{leads} to
\bea{
\p_{\theta_i}\mC(\theta_1,\theta_2)\aeq \int_0^\tau du \ \dbr{1}\mU^{\theta_1,\theta_2}(\tau,u)
\p_{\theta_i} \mL^{\theta_1,\theta_2}(u) \no\\
&\hs{5mm}\times \mU^{\theta_1,\theta_2}(u,0)\dke{\rho(0)} \ \ (i=1,2).
}
Using the above equation, $\mU^{0,0}(s,0)\dke{\rho(0)} = \dke{\rho(s)}$, and
\bea{
\dbr{1}\mU^{0,0}(s,0) \aeq \dbr{1} \la{br1_U}, 
}
we obtain \re{def_I_3}. 
\re{br1_U} is derived from $\dbr{1}\mL(s)=0$. 
The second derivative \re{def_I} consists of
\bea{
\int_0^\tau ds \ \tr_S\Big[\p_{\theta_1}\p_{\theta_2}\mL^{\theta_1,\theta_2}(s) \bv{\theta_1=0=\theta_2} \rho(s)\Big]=\f{1}{4}B(\tau) 
}
and 
\bea{
&\int_0^\tau ds \int_0^s du \  \tr_S\Big[\p_{\theta_{i^\pr}}\mL^{\theta_1,\theta_2}(s)\mU^{\theta_1,\theta_2}(s,u) \no\\
&\hs{5mm}\times\p_{\theta_i}\mL^{\theta_1,\theta_2}(u) \bv{\theta_1=0=\theta_2}\rho(u)\Big]
=I_i \ \ (i=1,2)
}
with $1^\pr \defe 2$ and $2^\pr\defe 1$. 
Note that 
\bea{
\mL_{i}(s) = \p_{\theta_i}\mL^{\theta_1,\theta_2}(s) \bv{\theta_1=0=\theta_2}.
}

$Q_+(t)$ can be written as
\bea{
Q_+(t) \aeq Q_a(t)+Q_b(t)\la{Q_ex} ,
}
where
\bea{
&\hs{-3mm}Q_b(t)  \no\\
&\hs{-3mm}\defe 4\int_0^t ds \int_0^s du \ \tr_S\Big[\mL_1(t) \mU(s,u) \mP(u)\mL_2(u)\rho(u)\Big] .
 }
 $q_a$ in \re{main_1} is defined by
 \bea{
 q_a(s) \defe \int_0^s du \ \mU(s,u) \mP(u)\mL_1(u) \rho(u).
 }
Because of
\bea{
(\mL \bu)\dg \aeq \mL \bu \dg ,\\
(\mL_i \bu)\dg \aeq \mL_{i^\pr} \bu \dg ,\\
(\mP\mL_i \bu)\dg \aeq \mP\mL_{i^\pr} \bu \dg,
} 
$Q_b=Q_a^\ast$ holds. 
Then, we obtain \re{Q_+_re}

\section{Quantum correction $Q_+$ in the long-time region} \la{A_correction}

The left and right eigenvalue equations of the Liouvillian $\mL$ are
\bea{
\mL\dke{\rho_n} \aeq \lm_n \dke{\rho_n } ,\la{eigen_R}\\
\dbr{l_n}\mL  \aeq \lm_n \dbr{l_n} \la{eigen_L}  . 
}
We set $\dbra l_m \dke{\rho_n }=\dl_{mn}$, 
$\lm_0=0$, and $\dbr{l_0}=\dbr{1}$. 
Then, $\rho_0(=:\rho^\RM{ss})$ is the steady state. 

If the initial state is the steady state, \re{Q_ex} becomes
\bea{
Q_+(\tau) \aeq 4 \Big( \int_0^\tau ds \int_0^s du \ \tr_S\Big[\mL_2 e^{(s-u)\mL} \mP \mL_1 \rho^\RM{ss}\Big]  \no\\
&\spa+\int_0^\tau ds \int_0^s du  \ \tr_S\Big[\mL_1 e^{(s-u)\mL} \mP \mL_2\rho^\RM{ss} \Big] \Big)
}
with 
\bea{
\mP \defe 1-\dke{\rho^\RM{ss}}\dbr{1}.
}
Using
\bea{
e^{(s-u)\mL} \aeq \sum_n e^{(s-u)\lm_n} \dke{\rho_n} \dbr{l_n},
}
we obtain
\bea{
&\hs{-3mm}\int_0^\tau ds \int_0^s du \ e^{(s-u)\mL} \mP \no\\
\aeq \sum_{n\ne 0} \int_0^\tau ds \int_0^s du \ e^{(s-u)\lm_n} \mP  \dke{\rho_n} \dbr{l_n} \no\\
\aeq -\tau \sum_{n\ne 0}\f{1}{\lm_n}\dke{\rho_n} \dbr{l_n} -\sum_{n\ne 0}\f{1}{\lm_n^2}\dke{\rho_n} \dbr{l_n} \no\\
&\spa+ \sum_{n\ne 0}\f{e^{\tau\lm_n}}{\lm_n^2}\dke{\rho_n} \dbr{l_n} \no\\
\aeqap -\tau \sum_{n\ne 0}\f{1}{\lm_n}\dke{\rho_n} \dbr{l_n} =-\tau \mR,
}
where 
\bea{
\mR \aeqd \sum_{n\ne 0}\f{1}{\lm_n}\dke{\rho_n} \dbr{l_n}=-\int_0^\infty dt \ e^{t\mL}\mP \la{def_R_0}
}
is the pseudo-inverse of the Liouvillian \cite{QTUR5, Flindt10, Nakajima15, Splettstoesser17, Nakajima17, Nakajima21} 
(the Drazin inverse \cite{Drazin} of $\mL$).
Thus, we obtain
\bea{
\hs{-2mm}Q_+(\tau) \approx -4\tau \Big( \tr_S\Big[\mL_2\mR \mL_1 \rho^\RM{ss}\Big]
 +\tr_S\Big[\mL_1\mR \mL_2\rho^\RM{ss}\Big] \Big)
}
in the long-time limit.
The expression of $Q_+$ in the long-time approximation is consistent with \rd{\re{dot_Q_+_long}}.

\section{Lower bound of $\mQ$} \la{s_Lower}

\re{F2} can be rewritten as
\bea{
\tl F=\sum_{m_0, \cdots,m_{N-1}=0}^M P^0(\{m_i \}) \Big[\p_\theta \ln P^\theta(\{m_i \}) \big\vert_{\theta=0} \Big]^2. \la{F1}
}
The score $\p_\theta \ln P^\theta(\{m_i \})\vert_{\theta=0}$ is given by
\bea{
\p_\theta \ln P^\theta(\{m_i \})\vert_{\theta=0} =b(\{m_i \})+d(\{m_i \})
}
where
\bea{
b(\{m_i \}) \aeqd \sum_{n=0}^{N-1}\dl_{m_n\ne 0} ,\\
d(\{m_i \}) \aeqd \sum_{n=0}^{N-1}\dl_{m_n,0} \ga_n(\{m_i \})\Dl t ,\\
\ga_n(\{m_i \}) \aeqd \f{\tr_S[\hat{\Om}_{m_{N-1}} \cdots \hat{\ga} \cdots  \hat{\Om}_{m_0} \rho(0)]}{P^0(\{m_i \})}  .
}
Here we introduce the convention that $\dl_{m\ne 0} \defe 1-\dl_{m,0}$ and $\hat{\Om}_m\bu \defe \Om_m^0 \bu (\Om_m^0 )\dg $. 
$\hat{\ga}$ is defined by \re{def_ga}.
Then  $\tl F$ \rd{becomes}
\bea{
\tl F \aeq \bra b(\{m_i \})^2 \ket +2 \bra b(\{m_i \})d(\{m_i \}) \ket+\bra d(\{m_i \})^2 \ket
}
where $\bra X(\{m_i \}) \ket \defe \sum_{m_0, \cdots,m_{N-1}=0}^M P^0(\{m_i \})X(\{m_i \})$. 
$b(\{m_i \})^2$ and $b(\{m_i \})d(\{m_i \})$ are calculated as
\rd{\bea{
b(\{m_i \})^2 \aeq b(\{m_i \}) +2\sum_{n=1}^{N-1}\sum_{l=0}^{n-1}\dl_{m_n\ne 0}\dl_{m_l\ne 0} ,\la{b^2}
}}
and 
\bea{
b(\{m_i \})d(\{m_i \}) \aeq \sum_{n=1}^{N-1}\sum_{l=0}^{n-1}\Big(\dl_{m_n\ne 0} \ga_l(\{m_i \})\dl_{m_l,0} \Dl t \no\\
&\spa+ \ga_n(\{m_i \})\dl_{m_n,0} \Dl t  \dl_{m_l\ne 0}\Big).
}
In \re{b^2}, the first and second terms of the right-hand side come from the contributions of the same times and different times respectively.
In the limit of $\Delta t \to 0$, trajectory averages of the above two equations become
\bea{
&\bra b(\{m_i \})^2 \ket = B(\tau) \no\\
&\spa+2\int_0^\tau ds\int_0^s du \ \tr_S[\hat{\Ga}(s) \mU(s,u)\hat{\Ga}(u) \rho(u)] ,\\
&\bra b(\{m_i \})d(\{m_i \}) \ket  \no\\
\aeq \int_0^\tau ds\int_0^s du \ \Big(\tr_S[\hat{\Ga}(s) \mU(s,u)\hat{\ga}(u) \rho(u)]  \no\\
&\spa+\tr_S[\hat{\ga}(s) \mU(s,u)\hat{\Ga}(u) \rho(u)]  \Big)
}
where
$\hat{\Ga}(t)\bu \defe \sum_k L_k \bu L_k\dg$.
Thus, we obtain
\bea{
&\hs{-3mm}\bra b(\{m_i \})^2 \ket +2 \bra b(\{m_i \})d(\{m_i \}) \ket \no\\
\aeq B(\tau)-2\int_0^\tau ds\int_0^s du \ \tr_S[\hat{\ga}(s) \mU(s,u)\hat{\ga}(u) \rho(u)]
}
and 
\bea{
&\mQ =\bra d(\{m_i \})^2 \ket \no\\
&\spa-2\int_0^\tau ds\int_0^s du \ \tr_S[\hat{\ga}(s) \mU(s,u)\hat{\ga}(u) \rho(u)] .
}
Here, $\bra d(\{m_i \})^2 \ket$ cannot be written as the double time integral because
$\ga_n(\{m_i \})$ in $d(\{m_i \}) $ depends on the entire sequence of a trajectory $\{m_i \}$.
Using $\bra d(\{m_i \})^2 \ket \ge \bra d(\{m_i \}) \ket^2$,  $\bra b(\{m_i \})+d(\{m_i \}) \ket=0 $, and $\bra b(\{m_i \}) \ket=B(\tau)$,
we obtain \re{Q_-}.

\section{Derivation\rd{s} of \re{dot_Q} \rd{and \re{dot_Q_-}} } \la{A_D}

\rd{We start from \re{mI_in_long}}. 
We suppose that the matrix representation of $\mL^{\theta_1,\theta_2}$ is block diagonalized, 
with one block $\tl \mL^{\theta_1,\theta_2}$ having eigenvalue $\lm(\theta_1, \theta_2)$.
The characteristic polynomial of $\tl \mL^{\theta_1,\theta_2}$ is given by
\bea{
\det(\tl \mL^{\theta_1,\theta_2}-\lm 1_{d}) =\sum_{n=0}^{d} a^{(n)}(\theta_1,\theta_2)\lm^n.
}
Here, $d$ is the dimension of $\tl \mL^{\theta_1,\theta_2}$ and 
$1_{d}$ is $d$-dimensional identity matrix. 
By differentiating $\sum_{n=0}^{d} a^{(n)}(\theta_1,\theta_2)\lm(\theta_1,\theta_2)^n=0$, we obtain
\bea{
a^{(0)}_i + a^{(1)}\lm_i = 0 \ \ (i=1,2) ,\la{lm_1}\\
a^{(0)}_{12}+a^{(1)}\lm_{12}+a^{(1)}_1\lm_2  +a^{(1)}_2\lm_1 +2a^{(2)}\lm_1 \lm_2 =0.\la{lm_2}
}
Here, $a^{(n)}\defe a^{(n)}(0,0)$,  $X_i \defe \p_{\theta_i} X \vert_{\theta_1=0=\theta_2}$, and  
$X_{12} \defe \p_{\theta_1}\p_{\theta_2} X \vert_{\theta_1=0=\theta_2}$.
$\lm_{12}$ is calculated from \re{lm_1} and \re{lm_2}. 
$\dot{Q}_+\defe \lim_{\tau \to \infty} Q_+/\tau$ is given by
\bea{
\dot{Q}_+\aeq 4\lm_{12}-\rd{\dot{B}_\RM{ss}} .\la{dot_Q_general}
}

For the system described by \re{Nu1}-\re{Nu3}, 
$\tl \mL^{\theta_1,\theta_2}$ is given by
\begin{widetext}
\bea{
&\tl \mL^{\theta_1,\theta_2} \no\\
\aeq \begin{pmatrix}
-\half(\Theta_1+\Theta_2)\ga n&i\Theta_2 \Om &-\Theta_1 \Om& \sqrt{\Theta_1\Theta_2}\ga(n+1)\\
i\Theta_2\Om& i\Theta_2\Dl -\half \Theta_2 \ga(n+1)-\half \Theta_1 \ga n &0&-i\Theta_1\Om\\
-i\Theta_1\Om &0&-i\Theta_1\Dl-\half \Theta_1 \ga(n+1)-\half \Theta_2 \ga n & i\Theta_2\Om \\
\sqrt{\Theta_1\Theta_2}\ga n &-i\Theta_1\Om&i\Theta_2 \Om&\mL_{44}
\end{pmatrix}
}
\end{widetext}
with $\mL_{44}\defe -\half(\Theta_1+\Theta_2)\ga (n+1) 
-i(\Theta_1-\Theta_2)\Dl$ and $\Theta_i\defe 1+\theta_i$. 
Using \re{lm_1}, \re{lm_2}, and \re{dot_Q_general}, we obtain \re{dot_Q}.

\rd{$Q_-$ can be rewritten as 
\bea{
Q_-(t) \aeq \f{\p^2}{\p \theta^2} \ln \tr_S \rho^\theta(t)\Bv{\theta=0}. \la{SLD_like}
}
Here, $\rho^\theta(t)$ is defined by
\bea{
\f{d}{dt}\rho^\theta(t) \aeq \mL^\theta \rho^\theta(t) ,\\
\mL^\theta \bu \aeqd (1+i \theta)\hat{\ga}\bu + \sum_k L_k \bu L_k\dg
}
with $\rho^\theta(0)=\rho(0)$. 
From \re{SLD_like}, we obtain 
\bea{
Q_-(\tau) \aeq \tau \f{\p^2}{\p \theta^2} \Lm(\theta)\Bv{\theta=0}+O(1)
}
where $\Lm(\theta)$ is the eigenvalue of $\mL^{\theta}$ which satisfies $\Lm(0)=0$. 
$\dot{Q}_-=\p^2\Lm(\theta)/\p \theta^2 \vert_{\theta=0} $ is calculated in the similar way as $\lm_{12}$. 
Then, we obtain \re{dot_Q_-}.
}

\section{Hasegawa's approach}  \la{A_Hasegawa}
 
We review Hasegawa's method and results \cite{H22}. 
We introduce a state
\bea{
\ke{\Phi(t)} \aeq U(t) \ke{\Phi(0)} 
}
with 
\bea{
\ke{\Phi(0)} \aeqd \ke{\tl \psi(0)} \otimes \ke{0} ,\\
U(t) \aeq \RM{T} \exp \Big[\int_0^t ds \ \Big\{-iH_S(s) \no\\
&+ \sum_k [L_k(s)\otimes \phi_k\dg (s)-L_k(s)\dg \otimes \phi_k (s)] \Big\} \Big].
}
The state $\ke{\Phi(t)}$ provides the solution of the GKSL equation \cite{H22}:
\bea{
\rho(t) \aeq \tr_{AB}[ \ke{\Phi(t)} \br{\Phi(t)} ] .\la{rho_Phi}
}
Using
$d\rho =\tr_{AB}\big[d(\ke{\Phi(t)} \br{\Phi(t)})\big]$, 
\bea{
d(\ke{\Phi(t)} \br{\Phi(t)})\aeq d\ke{\Phi(t)} \br{\Phi(t)} +\ke{\Phi(t)} d\br{\Phi(t)} \no\\
&\spa+d\ke{\Phi(t)} d\br{\Phi(t)} ,
}
and 
\bea{
d\ke{\Phi(t)} \aeq \Big(\big( -iH_S-\f{1}{2}\sum_k L_k\dg L_k\big) dt 
+ \sum_k L_k d\phi_k\dg \no\\
&\spa+\half \sum_{k,l} L_kL_l d\phi_k\dg d\phi_l\dg\Big)\ke{\Phi(t)}, \la{dPhi}
}
we obtain the GKSL equation \re{QME}. 
Here, $d\phi_k\dg=\int_t^{t+dt}ds \ \phi_k\dg(s)$. 

If we put
\bea{
\rho_\tau(s;t_1,t_2) \aeqd \tr_{AB}[ \ke{\Psi_\tau(s;t_1)} \br{\Psi_\tau(s;t_2)} ],
}
we obtain $\rho_\tau(\tau;t,t) =\rho(t)$ \cite{H22}.
The time evolution equation of $\rho_\tau(s;t_1,t_2)$ is given by
\bea{
\f{\p \rho_\tau(s;t_1,t_2) }{\p s} \aeq \mL_\tau(s;t_1,t_2) \rho_\tau(s;t_1,t_2) \la{two_1}
}
with 
\bea{
&\hs{-4mm}\mL_\tau(s;t_1,t_2) \bu \no\\
\aeq -i\f{t_1}{\tau} H_S\Big(\f{t_1}{\tau}s\Big) \bu +\bu i \f{t_2}{\tau} H_S\Big(\f{t_2}{\tau}s\Big) \no\\
&\spa +\sqrt{\f{t_1}{\tau}}\sqrt{\f{t_2}{\tau}}\sum_k L_k\Big(\f{t_1}{\tau}s\Big)\bu  L_k\Big(\f{t_2}{\tau}s\Big)\dg \no\\
&\spa -\half \sum_k \Big[\f{t_1}{\tau}L_k\Big(\f{t_1}{\tau}s\Big)\dg L_k\Big(\f{t_1}{\tau}s\Big) \bu \no\\
&\hs{12mm} +\bu \f{t_2}{\tau}L_k\Big(\f{t_2}{\tau}s\Big)\dg L_k\Big(\f{t_2}{\tau}s\Big) \Big]. \la{two_1L}
}
\re{two_1} is a two-sided GKSL equation. 
Because of 
\bea{
\mL_\tau(s;t,t) \aeq \f{t}{\tau} \mL\Big( \f{t}{\tau}s \Big),
}
we obtain
\bea{
\rho_\tau(s;t,t) \aeq \rho\Big(\f{t}{\tau}s\Big). \la{tau_to_t}
}

$\mJ(t)$ defined by \re{def_mJ} is given by
\bea{
\mJ(t) \aeq  4[\p_{t_1}\p_{t_2}C(t_1,t_2) \no\\
&\spa-\p_{t_1}C(t_1,t_2)\p_{t_2}C(t_1,t_2) \Big]\Bv{t_1=t=t_2} \la{def_mI} 
}
with $C(t_1,t_2) \defe \tr_S \rho_\tau(\tau;t_1,t_2) $.
$\mJ(t)$ is independent from $\tau$. 

From \re{MTR}, Hasegawa \cite{H22} showed a KUR: 
\bea{
\f{\tau^2(\p_\tau \bra \mC \ket_\tau)^2}{\bra \mC^2 \ket_\tau-\bra \mC \ket_\tau^2} 
\le \mB(\tau). \la{S58} 
}
Here, 
\bea{
\mB(t) \defe t^2 \mJ(t)
}
is \co{the quantum generalization of the dynamical activity}. 
$\bra X \ket_\tau \defe \br{\Psi_{\tau}(\tau;\tau)} X  \ke{\Psi_{\tau}(\tau;\tau)}=\br{\Phi(\tau)} X\ke{\Phi(\tau)}$
and $\mC$ is an operator of the field system which describes a time-integrated counting observable: 
$\mC$ counts and weights jump events in a quantum trajectory of the system $S$. 
In Ref.\cite{H22}, Hasegawa showed a KUR for more general operators of the field system. 

If $H_S$ and $L_k$ are time-independent, \re{two_1L} is obtained by replacing  
$1+\theta_i$ by $t_i/\tau $ ($i=1,2$) in \re{two_2L}. 
In this case, $\mB(\tau)$ in Ref.\cite{H22} is identical to $\mI$ in Ref.\cite{Vu22}:
\bea{
\mB(\tau) = \mI.
}
\re{S58} is consistent with \rd{\re{KUR_H}}.

\section{Quantum dot}  \la{A_QD}

We consider the quantum dot \re{QME_QD}. 
The state of the system can be written as $\tl \rho(t)=\half(1+\bm{r}(t)\cdot \bm{\sig})$.
Here, $\bm{\sig}=(\sig_x,\sig_y,\sig_z)$, $\sig_i$ is the Pauli matrix, and  $\bm{r}(t)=(x,y,z)$ is the Bloch vector. 
The equation of the motion of the Bloch vector is given by
\bea{
\f{d}{dt}x\aeq -\half \ga x \com
\f{d}{dt}y = -\half \ga y ,\\
\f{d}{dt}z\aeq -\ga (z-[1-2f(\ep)]).
}
The dynamical activity is given by
\bea{
B(t)=\int_0^t ds \ \f{\ga}{2}\big(1+[2f(\ep)-1]z(s)\big).
} 
We put $\rho_i \defe \tl \rho(t_i)$. 
If the eigenvalues of $\ka \defe \sqrt{\rho_1} \rho_2 \sqrt{\rho_1}$ are $\lm_1$ and $\lm_2$, 
the fidelity is given by $F(\rho_1, \rho_2)=\sqrt{\lm_1}+ \sqrt{\lm_2}$. 
Then, 
\bea{
[F(\rho_1, \rho_2)]^2\aeq \tr_S(\ka)+2\sqrt{\det(\ka)} \no\\
\aeq \tr_S(\rho_1\rho_2)+2\sqrt{\det(\rho_1) \det(\rho_2)}.
}
This leads to
\bea{
&F(\tl \rho(t_1), \tl \rho(t_2)) \no\\
\aeq \sqrt{\f{1+\bm{r}(t_1)\cdot\bm{r}(t_2)+\sqrt{[1-\bm{r}(t_1)^2][1-\bm{r}(t_2)^2]}}{2}}.
}
The Bures angle is given by $D(\tl \rho(t_1), \tl \rho(t_2))=\cos^{-1}F(\tl \rho(t_1), \tl \rho(t_2))$.
The trace distance is given by 
\bea{
T(\tl \rho(t_1), \tl \rho(t_2)) = \half \abs{\bm{r}(t_1)-\bm{r}(t_2)}.
} 
Here, $\abs{\bm{x}}=\sqrt{\bm{x}^2}=\sqrt{\bm{x}\cdot \bm{x}}$.

\end{document}